\documentclass[groupaddress,prb,twocolumn]{revtex4}

\usepackage{graphicx}
\usepackage{amssymb}

\begin{document}

\title{Active metamaterials: sign of refraction index and gain-assisted dispersion management }

\author{Alexander A. Govyadinov}
\affiliation{Physics Department, Oregon State University, 301 Weniger Hall, Corvallis, OR 97331, USA}
\author{M. A. Noginov}
\email{mnoginov@nsu.edu}
\affiliation{Center for Material Research, Norfolk State University, Norfolk, VA 23504, USA}
\author{Viktor A. Podolskiy}
\email{viktor.podolskiy@physics.oregonstate.edu}
\affiliation{Physics Department, Oregon State University, 301 Weniger Hall, Corvallis, OR 97331, USA}


\begin{abstract}
We derive an approach to define the causal direction of the wavevector of modes in optical metamaterials, which in turn, determines signs of refractive index and impedance as a function of {\it real and imaginary} parts of dielectric permittivity and magnetic permeability. We use the developed technique to demonstrate that the interplay between resonant response of constituents of metamaterials can be used to achieve efficient dispersion management. Finally we demonstrate broadband dispersion-less index and impedance matching in active nanowire-based negative index materials. Our work opens new practical applications of negative index composites for broadband lensing, imaging, and pulse-routing.
\end{abstract}

\maketitle

Recent research in the area of negative index materials (NIMs) \cite{NIMorigin} has resulted in a number of exciting applications, including super-imaging\cite{superimaging,hyperlens} and subwavelength light compression\cite{compressors}. However, the majority of these applications suffer from substantial material absorption\cite{NIMlosses} and frequency dispersion of NIM composites. Optical gain has been suggested to minimize, and potentially eliminate absorption losses\cite{gainNIM,gainSPP,gainLoc}. While effect of gain on propagation length of optical signals is straightforward, its effect on material dispersion has not been completely understood. Furthermore, the very question of sign of refractive index in active metamaterials is somewhat controversial\cite{lakhtakiaCom,wise}. In this Letter we present a universal approach for imposing causality in active and passive materials, and use this technique to analyze the perspectives of gain-assisted dispersion management beyond loss compensation. 

Our results can be used to determine the sign of refractive index of active or passive media, as well as in a number of analytical and/or numerical solutions of Maxwell equations relying on plane-wave representations. Our technique is illustrated using an example of nanowire-based NIM structure, originally proposed in\cite{nanowires} and experimentally realized in\cite{nanowiresExp}. It is shown that frequency independent (negative) index and impedance can be achieved in the same metamaterial with position-dependent gain in weak gain regime. A combination of broadband impedance and refractive index has a potential to open new exciting applications of dispersion-managed NIMs, in broadband optical processing, packet routing, and non-reflective lensing.

Index of refraction $n_{\rm ph}$ is one of the most fundamental optical properties of the media. Its magnitude relates the magnitude of wavevector $\vec{k}$ of plane electromagnetic wave to frequency $\omega$:  $|n_{\rm ph}|= {|\vec{k}|c}/{\omega}$, and thus describes the phase velocities of waves in the material\cite{Jackson}. In particular, $n_{\rm ph}$ enters the equations for reflectivity, Doppler effect, and some nonlinear phenomena. Apart from $n_{\rm ph}$, reflectivity of a material also depends on its impedance $Z$. For an isotropic material with (complex) dielectric permittivity $\epsilon$ and magnetic permeability $\mu$, $n_{\rm ph}$ and $Z$ are calculated via\cite{Jackson}:
\begin{eqnarray}
\label{eqNeff}
  n_{\rm ph} =\pm \sqrt{\epsilon \mu}\\
\label{eqZeff}
  Z = \mu/n_{\rm ph} = \pm \sqrt{\mu/\epsilon},
\end{eqnarray}
Note that while the magnitude of $n_{\rm ph}$ and $Z$ are completely determined by material paramaters ($\epsilon$ and $\mu$)\cite{Jackson}, their signs have recently instigated some controversy\cite{lakhtakiaCom,lakhtakiaNIM,efrosSign,wise}, which can be traced to different treatments of causality principle. Moreover, the authors of \cite{efrosSign} suggest that Maxwell equations can be solved correctly regardless the selection of sign of refractive index. Such a freedom of choice, however is accompanied by a requirement to adjust the signs in equations describing phase velocity-related phenomena, e.g. Snell's law\cite{efrosSign}, and still require imposing causality (identical to \cite{lakhtakiaNIM}) when solving Maxwell equations. 

From a mathematical standpoint, imposing causality principle is equivalent to selecting the sign of the wavevector of a plane wave propagating away from the source. Here we assume that such a propagation takes place along the positive $z$ direction and therefore focus on the $k_z$ component of the wavevector. The authors of \cite{lakhtakiaCom,lakhtakiaNIM,efrosSign} propose to select the sign of $k_z$, enforcing positive direction of Poynting vector (associated with energy flux). The authors of \cite{wise} suggest that causality requires exponential decay ($k_z ''>0$\cite{footnote1}) of waves propagating inside passive materials and exponential growth ($k_z ''<0$) of waves propagating inside active media. 

While all causality requirements discussed above\cite{lakhtakiaNIM,lakhtakiaCom,wise,efrosSign} coincide for the case of passive materials, they are not directly applicable for active media and are therefore not universal. Indeed, enforcing the sign of energy flux is physical only in transparent materials. Materials with opposite signs of $\epsilon$ and $\mu$ (known as single-negative materials\cite{compressors,Jackson}) reflect the majority of incident radiation. Enforcing decay/growth of field based solely on passive/active state of material yields nonphysical results (such as abrupt disappearance of surface plasmon polariton waves) when media undergoes smooth transition from low-loss to low-gain state\cite{noginovPress}.

To resolve the above controversy\cite{wise,lakhtakiaCom}, we propose to simultaneously consider ``eigen transparency'' of the material (its transparency in the absence of losses and gain, $\epsilon''=\mu''=0$) along with absorption (or gain) state of the material. Clearly, electromagnetic radiation should decay inside all passive media\cite{Jackson}. It should also grow inside {\it transparent} (double-negative or double-positive) active materials. Non-transparent (single-negative) materials do not support propagating modes, and thus should reflect all incident light. Energy can penetrate these structures only in the form of exponentially decaying waves\cite{footnoteDecay}. Since decay/growth of EM waves can be related to the sign of imaginary part of refraction index, our arguments, summarized in Table I, provide complete solution to the problem of selection of direction of the wavevector of plane waves. For isotropic media, the developed technique also provides a solution to selection of the sign of $n_{\rm ph}'$, which should be identical to that of $k_z'$, yielding ``conventional'' Snell's law\cite{NIMorigin}. 

For passive media, our results agree with those of\cite{efrosSign,lakhtakiaNIM}, and with \cite{wise}, relying on the pre-selected branch cut in the complex plane when calculating the square root in Eq.(\ref{eqNeff}). We note however, that Table I cannot be reduced to such a cut. Indeed, an optical material can fall into one of the four cases: it has either negative ($n_{\rm ph}'< 0$)or positive ($n_{\rm ph}'> 0$) refractive index, and it attenuates ($n_{\rm ph}''\geq 0$) or amplifies ($n_{\rm ph}''<0$) incoming radiation\cite{footnote1}. Selection of any single complex plane cut in Eq. (\ref{eqNeff}) would immediately limit the number of possible $\{n_{\rm ph}',n_{\rm ph}''\}$ combinations to two, and therefore in general is not correct.

While it is impossible to determine a universal single cut in the complex plane for $n_{\rm ph}$, requirements of Table I can be formally satisfied by the following procedure: starting from material parameters $\epsilon$ and $\mu$, one first calculates $\sqrt{\epsilon}$ and $\sqrt{\mu}$, cutting the complex plane {\it along negative imaginary axis} (see \cite{footnoteDecay} and discussion in Ref.\cite{noginovPress} on implications of different selection of signs of $k_z''$ in non-magnetic media). Refraction index and impedance are then calculated as $n_{\rm ph}= \sqrt{\epsilon}\cdot \sqrt{\mu }$; $Z= \sqrt{\mu}/ \sqrt{\epsilon }$

\begin{table}
  \centering
  \begin{tabular}{|c|c|c|}
    \hline
    Transparency & Gain/Loss & Wave Growth/Decay \\
    $\epsilon' \cdot \mu'$ & $|\epsilon| \epsilon'' +|\mu| \mu''$ & $k_z''$ \\
    \hline
    $+$ & $+$ & $+$ \\
    $+$ & $-$ & $-$ \\
    $-$ & any & $+$ \\
    \hline
  \end{tabular}
  \caption{The direction of wavevector (and thus the sign of refractive index) in optical material is related to the interplay between transparency and gain/loss state of the media. The table summarizes this dependence. First column represents the transparency state of the material, determined by the sign of product $\epsilon' \mu'$; The sign of $|\epsilon| \epsilon'' +|\mu| \mu''$ (second column) determines whether material is passive ($|\epsilon| \epsilon'' +|\mu| \mu''>0$) or active ($|\epsilon| \epsilon'' +|\mu| \mu''<0$). The sign of the refractive index is selected to satisfy the requirement for wave attenuation ($k_z''>0$) or growth ($k_z''<0$) (third column) }\label{tblImN}
\end{table}

We note that the above procedure can be generalized to other classes of materials and excitation waves. For example, similar transparency/active state arguments can be employed to find the ``handedness'' (relationship between directions of $\vec{E}$, $\vec{H}$, and $\vec{k}$) of modes in active anisotropy-based waveguides\cite{podolskiyPRB}, as well as the handedness of waves in active media excited by evanescent radiation\cite{noginovPress,plotz} (the latter case is typically realized in gain-assisted compensation of losses of surface plasmon polaritons\cite{gainSPP}).

\begin{figure}[b]
  \includegraphics[width=8.5cm]{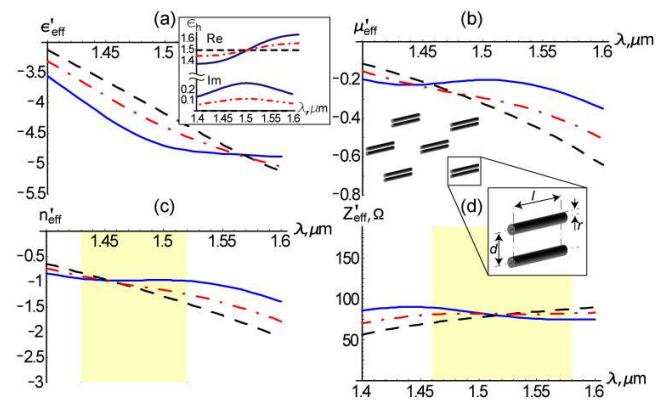}\\
  \caption{Effective permittivity (a), permeability (b), refractive index (c) and impedance (d) of the passive nanowire NIMs described in text and schematically shown in inset in (b); solid lines corresponds to $A=0.04$, dashed to $A=0$, dash-dotted to $A=0.0175$. Inset in (a) shows real and imaginary parts of permittivity of host material}
\label{fig1}
\end{figure}

We now employ the developed technique to analyze the gain-assisted dispersion management in active negative-index metamaterials. To illustrate our approach, we select nanowire-based optical NIM system\cite{nanowires,nanowiresExp}. The meta-atom of this composite comprises two plasmonic nanowires, schematically shown in inset of Fig.\ref{fig1}b. As described in detail in \cite{nanowires}, dielectric and magnetic response of such a metamaterial can be related to symmetric and anti-symmetric combinations of surface plasmon polaritons excited in nanowire pairs. In the limit of small concentration of nanowires $p$, effective dielectric permittivity and magnetic permeability of such a mix can be qualitatively described by \cite{nanowires}:
\begin{eqnarray}
 \label{eqEpsEff}
  \epsilon_{\rm eff}=\epsilon_h + \frac{4 p r}{d} \frac{{\rm f}(\Delta)\epsilon_m}{1+\frac{4{\rm f}(\Delta)\epsilon_m r^2}{l^2} \ln \left(1+\frac{\epsilon_h l}{2r}\right) \cos\Omega}, \\
 \label{eqMuEff}
  \mu_{\rm eff}=1+\frac{12pl^2 C_2 k^2 d^2}{rd} \frac{2 \tan(gl/2)-gl}{(gl)^3},
\end{eqnarray}
where, $r$, $l$, and $d$ correspond to nanowire radius, length, and separation between two wires, and remaining parameters are given by:
$\Omega^2= {k^2 l^2} ( {\ln[l/2r]+i\sqrt{\epsilon_h}kl/2})/(4\;{\ln\left[1+{\epsilon_h l}/{2r}\right]})$,
$C_2=\epsilon_h/(4 \ln[d/r])$,
$g^2=k^2\epsilon_h \left[1+ {i}/({2\Delta^2 \: {\rm f}[\Delta] \ln[d/r]})\right]$,
$\Delta=k r \sqrt{-i\epsilon_m}$,
${\rm f}(\Delta)=(1-i)
{J_1 [(1+i)\Delta]}/(\Delta {J_0 [(1+i)\Delta]})$,
with $k=2 \pi/\lambda=\omega/c$,
$\lambda$ is being wavelength in the vacuum, and $\epsilon_m$ and $\epsilon_h$ being permittivities of nanowires and host materials. Here we assume that $\epsilon_m$ of silver wires is described by the Drude model \cite{Kittel}, and further assume that host dielectric consists of a polymer ($\epsilon_0 \simeq 1.5$) doped with quantum dots\cite{qdots}, qualitatively described by Lorentz model:
\begin{equation}\label{eqHost}
    \epsilon_h=1.5 + \frac{A \omega_0^2}{\omega_0^2-\omega^2-i \omega \gamma},
\end{equation}
where $\omega_0$ is the resonant frequency, $\gamma$ is the damping constant, and $A$ is the macroscopic analog of Lorentz oscillator strength, which formally describes gain in the system and can be related to the concentration of quantum dots and the fraction of quantum dots in excited state. $A>0$ corresponds to lossy materials; $A=0$ represents the case when the number of excited quantum dots is equal to the number of dots in the ground state; $A<0$ corresponds to the inverted (gain) regime\cite{footnoteGain}. The permittivity of the host medium and corresponding permittivity of the NIM system for different pump rates are shown in inset of Fig.\ref{fig1}a.

Fig.\ref{fig1}a illustrates perspectives of dispersion management in {\it lossy} ($A \geq 0$) nanowire composites with $p=0.1$, $r=25nm$, $l=700nm$, $d=120nm$, and $\lambda_0=2 \pi c/\omega_0 = 1.5\mu m$ $\gamma=0.628 \mu m^{-1}$. It is clearly seen that dispersion of host media can completely compensate the dispersion of refractive index and impedance of the NIM system. Note, however, that due to their different dependence on $\epsilon$ and $\mu$, broadband refractive index and broadband impedance are realized at different values of oscillator strength $A$ in the single-oscillator model assumed here. We suggest that benefits of impedance-matching can be combined with benefits of index-matching in the same system where $A$ is (adiabatically) changed from $A\simeq 0.0175$ corresponding to $\partial Z/\partial\omega=0$ at the interface to $A\simeq 0.04$ corresponding to $\partial n_{\rm ph}/\partial \omega=0$ in the core of the system. In quantum dot materials, spatial change of $A$ can be easily achieved by changing quantum dot doping or external pumping rate.

Although passive host ($A>0$) does yield broadband frequency-independent $n_{\rm ph}$ and $Z$, it also increases total absorption in NIM structure, further limiting its practical size to $\protect {\lesssim 1\cdots }10 \mu m$\cite{NIMlosses}. Active quantum dots, on the other hand can simultaneously reduce absorption in the system and provide versatile gain-assisted dispersion management. Note, that such a modulation of $n_{\rm ph}$ or $Z$ does not require full compensation of propagation losses.

\begin{figure}
  \includegraphics[width=8.5cm]{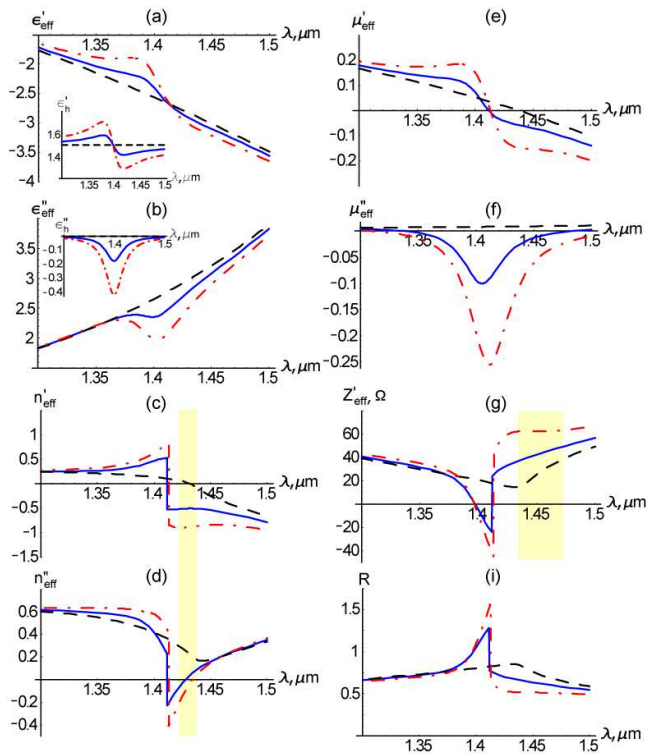}\\
  \caption{ Real and imaginary parts of the effective permittivity (a,b), permeability (e,f), refractive index (c,d), real part of impedance (g) of of active nanowire NIMs, and reflection from the semi-infinite slab of this system (i). Solid, dashed, and dash-dotted curves correspond to $A=-5.05\cdot10^{-3}$, $A=0$ and $A=-0.012$ respectively. Insets in (a) and (b) show the real and imaginary parts of permittivity of host material respectively.}\label{fig2}
\end{figure}

Gain-assisted dispersion management in active nanowire composites with $A<0$, $\lambda_0=1.4\mu m$, $\gamma=0.129 \mu m^{-1}$, $p=0.09$, $r=25nm$, $l=720nm$, and $d=120nm$ are shown in Fig.\ref{fig2}. Note that refraction index, impedance, as well as reflectivity between vacuum and nanowire metamaterial are continuous when material switches between active and passive states. In contrast to this behavior, transition between transparent and ``metallic'' regimes yields a discontinuity in reflectivity (the discontinuity disappears when thickness of gain region is finite). This discontinuity\cite{plotz} is accompanied by enhanced reflection ($R>1$), and has a physical origin similar to the one of enhanced reflectivity reported in \cite{plotz,noginovPress} for gain media excited by evanescent waves in total internal reflection geometry. 

To conclude, we developed a universal approach to determine the sign of refraction index and impedance in active and passive media. We have further utilized this approach to demonstrate versatile dispersion management, achieving $\partial n_{\rm ph}/\partial\omega=0$ and $\partial Z/\partial\omega=0$ regimes in nanowire-based NIM system with bandwidth equivalent to pico-second optical pulses. The developed technique can be readily utilized to determine sign of refraction index in different classes of materials and structures, including split-ring- and fish-scale geometries, waveguides-based and anisotropy-based NIMs, and can be used to optimize the dispersion of these structures for various photonic applications. Furthermore, a combination of several dopants with tailored gain/absorption spectra can be used to engineer a metamaterial having $\partial n_{\rm ph}/\partial\omega=0$ and $\partial Z/\partial\omega=0$ in the same frequency range. We note that handedness and dispersion of modes in $nm$-thick metamaterials will be strongly affected not only by dispersion of their constituents, but also by overall geometry\cite{vpvgPRL}. In general, NIM structures with smaller losses \cite{lowloss,podolskiyPRB} will exhibit weaker dispersion, potentially increasing operating bandwidth, and simultaneously reducing the gain values, required to achieve efficient dispersion management. In particular we expect gain $\lesssim 100cm^{-1}$ to be sufficient in nonmagnetic anisotropy-based NIMs \cite{podolskiyPRB}.

The authors acknowledge fruitful discussions with A.L. Efros. This research has been partially supported by Petroleum Research Fund (ACS), Army Research Office, Office of Naval Research, NSF PREM grant \# DMR 0611430, NSF NCN grant \# EEC-0228390, NSF CREST grant \# HRD 0317722, and NASA URC grant \# NCC3-1035.



\begin{thebibliography}{}
\bibitem{footnoteDecay}Excitation of waves in active singe-negative materials is equivalent to excitation of waves in conventional active media (dye solution) by evanescent waves, commonly realized in SPP experiments. As shown in Ref.\cite{noginovPress}, selection of exponentially growing solution in active singe-negative materials will lead to unphysical results (such as disappearance of SPP mode)
\bibitem{footnoteGain}We note that single-oscillator model, although it provides correct qualitative description of material properties, does not accurately describe active media, where emission and absorption bands typically occur at different frequencies
\end{thebibliography}
\end{document}